\documentclass[aps,PRB,reprint,longbibliography,superscriptaddress]{revtex4-2}
\usepackage{amsmath,amssymb}
\usepackage{graphicx}
\graphicspath{ {figure1} }
\usepackage{dcolumn}
\usepackage{bbold}
\usepackage[hidelinks]{hyperref}
\hypersetup{
  colorlinks,
  citecolor=magenta,
  linkcolor=blue,
  urlcolor=blue}
\usepackage{verbatim,color,xcolor}

\newcommand{\HB}{H_{\text{QSH}}}
\newcommand{\HS}{H_{\text{SC}}}
\newcommand{\gp}{g_{\parallel}}

\begin{document}

\title{Quantum spin Hall insulator in proximity with a superconductor: Transition to the Fulde-Ferrell-Larkin-Ovchinnikov state driven by a Zeeman field}

\author{Suman Jyoti De}
\affiliation{Harish-Chandra Research Institute, A CI of Homi Bhabha National Institute, Chhatnag  Road, Jhunsi, Prayagraj 211019, India}
\author {Udit Khanna} 
\affiliation{Department of Physics, Bar-Ilan University, Ramat Gan 52900, Israel}
\author{Sumathi Rao}
\affiliation{International Centre for Theoretical Sciences (ICTS—TIFR), Shivakote, Hesaraghatta Hobli, Bangalore 560089, India}
\author{Sourin Das}
\affiliation{Department of Physical Sciences, IISER Kolkata, Mohanpur, West Bengal 741246, India}

\begin{abstract}
We investigate the effects of introducing a boost (a Zeeman field parallel to the spin quantization axis) at the proximitized helical edge of a two-dimensional (2D) quantum spin Hall insulator. 
Our self-consistent analysis finds that a Fulde-Ferrell-Larkin-Ovchinnikov (FFLO) superconducting phase may emerge at the edge when the boost is larger than a critical value tied to the induced pairing gap. 
A non-trivial consequence of retaining the 2D bulk in the model is that this boundary FFLO state supports a finite magnetization as well as finite current (flowing along the edge). 
This has implications for a proper treatment of the ultra-violet cutoff in analyses employing the effective one-dimensional (1D) helical edge model. 
Our results may be contrasted with previous studies of such 1D models, which found that the FFLO phase either does not appear for any value of the boost (in non-self-consistent calculations), or that it self-consistently appears even for infinitesimal boost, but carries no current and magnetization. 
\end{abstract}

\maketitle

\section{Introduction}
One dimensional (1D) topological superconductors support exotic zero energy Majorana modes localized at their boundaries~\cite{Kitaev_2001,Zhang_2011_rev,Beenakker2013_rev}. 
The non-Abelian nature of these modes, and the ensuing potential applications in quantum computation~\cite{Kitaev_2003,Nayak_RMP_2008}, have generated considerable interest in exploring different platforms that may realize topological superconductivity~\cite{Sau2010,Oreg2010,Ando_Review_2017,Stern_Review_2021,Kristian2023}. 
The 1D helical edge modes~\cite{Wu2006,Xu2006} of two-dimensional (2D) quantum spin Hall (QSH) insulators are particularly promising in this context~\cite{Kane2005-1,Kane2005-2,Bernevig2006,Konig2007,Liu2008}. 
Topological superconductivity is expected to be induced at the helical edge directly through the proximity effect of conventional $s$-wave superconductors~\cite{FuKane2008,FuKanePRB2009}. 
Moreover, the inherently 2D nature of the setup allows for flexibility in device design that may be exploited to gain different advantages, including the possibility of immunity from disorder-induced backscattering~\cite{Vivek_chiral,Li2013}, which remains a formidable challenge in 1D nanowire based setups~\cite{Frolov2021,DasSarma2021A,DasSarma2021B,DasSarma2022}. 

Zeeman fields parallel and perpendicular to the spin-quantization axis of the helical edge modes offer a tunable 2D parameter space to probe the nature of the induced superconductivity and phase transitions away from it. 
A perpendicular field tends to open a band gap at the edge and drives a transition from the topological superconductor to a trivial insulator~\cite{Alicea2012}. 
On the other hand, in the absence of superconductivity, a parallel Zeeman field shifts (or boosts) the Dirac node away from $k = 0$ towards finite momentum, thereby inducing a persistent charge current at the helical edge~\cite{Vivek_chiral,Meyer2015}. 
In the rest of this Letter, we shall refer to this parallel Zeeman field as `boost'. 
The combined effects of boost and induced superconductivity at the helical edge is expected to lead to exciting phenomena, such as anomalous Josephson relations~\cite{Meyer2015} and the possibility of gapless superconductivity~\cite{Vivek_chiral}.  
However, in most of these works superconductivity is treated phenomenologically by adding a constant order parameter to the effective 1D Hamiltonian of the helical edge. 
This neglects the possibility for spatial variation of the order parameter~\cite{Annica2011} and the effect of the boost on the nature of superconducting state.

\begin{figure}[t]
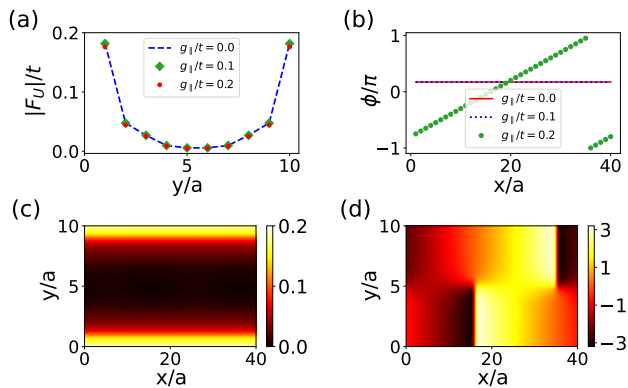

    \centering
    \includegraphics[width=\columnwidth]{{{Magnitude_and_phase_Delta_plot_2.pdf}}}
    \caption{Evolution of the local order parameter $F_{U}$ with boost $\gp$.  
    (a) The variation of the magnitude of $F_{U}$ (at $x = 20 a$) with $y$ for different $\gp$. 
    Clearly, the pairing amplitude decays exponentially in the bulk (as expected). 
    For all cases considered here, the magnitude remains uniform along the edge (along $x$). 
    (b) The variation of the phase of $F_{U}$ at the top edge ($y = 10 a$) with $x$ for different $\gp$. 
    For low values of $\gp$, the phase remains uniform along the edge, pointing to a BCS-like superconducting state. 
    For sufficiently large $\gp$ [for $0.2 t$ in (b)], the phase winds from $0$ to $2\pi$ once over the length of the edge (from $x = 0$ to $x = 40a$), indicating the emergence of an FFLO state. 
    (c, d) Spatial distribution of the magnitude [in (c)] and phase [in (d)] of $F_{U}$ in the FFLO phase (for $\gp = 0.2 t$). 
    Here, we used $U = 3 t$, $M_{0} = -2 t$ and $B/a^2 = -t$.}
    \label{fig:1}
\end{figure}

In this Letter, we study the interplay of edge superconductivity with the applied boost and the topological 2D band structure of the QSH insulator through a self-consistent analysis. 
Our calculation not only accounts fully for the effect of boost on both the amplitude and phase of the order parameter, but also correctly treats the role of the lattice and the finite bandwidth of the 2D band structure. 
The results suggest that the induced pairing is BCS-like, with a uniform order parameter, for small values of boost whereas a Fulde-Ferrell-Larkin-Ovchinnikov (FFLO) phase~\cite{FFLO-1,FFLO-2} emerges if the boost is larger than a certain critical value (see Fig.~\ref{fig:1}). 
This transition occurs when the boost is comparable to the induced pairing amplitude (in the absence of boost). 

Applying a boost to the helical edge, not only shifts the Dirac cone to finite momentum, but also generates a charge imbalance between the spin-up and down channels, or equivalently left and right movers due to rearrangement of states at the bottom of the band. 
Hence, this imbalance leads to a finite magnetization and current at the edge, and we find that (in a self-consistent analysis) this affects the nature of the induced superconductivity. 
However, such effects are tricky to take into account in effective 1D models of the edge, as the exact charge of the ground state depends on how the ultra-violet cutoff is implemented. 
Indeed previous works, performing self-consistent analyses within such 1D models, found that an FFLO phase may emerge in response to even an infinitesimal boost~\cite{Meyer2015} and that this ground state carries no net current or magnetization, an indication of no charge imbalance. 
By contrast, we are able to account for these effects properly by employing a 2D lattice model, and find that the FFLO ground state does carry signatures of the charge imbalance, namely current and magnetization (see Fig.~\ref{fig:Cur_Mag_SC}). 
The results show that the spatially varying order parameter (in the FFLO state) tends to reduce the current induced by the boost in the non-superconducting helical edge, but is unable to cancel it completely within our setup. 

\section{Model Hamiltonian and Analysis}

We employ the standard BHZ model~\cite{Bernevig2006} to describe a 2D QSH insulator, assuming that the spin-quantization axis of the helical edge is along the out-of-plane direction~\cite{Weithofer2013}. 
Then the continuum Hamiltonian of the boosted QSH insulator is,
\begin{align}
  \HB = \sum_{\bf k} \Psi_{\bf k}^\dagger \Big[v_F &\big(k_x \sigma_x \tau_z + k_y \sigma_y \mathbb{1}_{\tau} \big)  \nonumber \\
  &+ \big( M_0 - Bk^2 \big) + \gp \mathbb{1}_\sigma \tau_z \Big] \psi_{\bf k}. 
  \label{H_BHZ}
\end{align}
Here $\tau$ and $\sigma$ are Pauli matrices acting in the spin (${\uparrow},{\downarrow}$) and orbital ($A,B$) subspace respectively and we used the basis, $\psi_{\bf k} = (c_{{\bf k}A\uparrow}, c_{{\bf k}A\downarrow}, c_{{\bf k}B\uparrow}, c_{{\bf k}B\downarrow})^T$. 
$\HB$ describes a topologically non-trivial phase for $B < 0$ and $M_0 < 0$. 
$\gp$ in the last term is the boost applied uniformly throughout the setup. 
We choose $\gp \ll 2|M_{0}|$ (the bulk band gap) so that the boost only affects the edges while bulk remains inert.  
Throughout this work, the chemical potential is set to zero, implying that the Fermi energy lies at the Dirac node of the helical edge. 

In order to perform the calculations, we discretize $\HB$ over a square lattice (with lattice constant $a$) using standard methods~\cite{Vivek_chiral}. 
The largest nearest neighbor hopping integral is $t = v_{F} / a$. 
For all results presented here, the system size was $40a$ ($10a$) along the $x$ ($y$) direction. 
We imposed open (periodic) boundary conditions along the $y$ ($x$) direction. 
The periodic boundary conditions along $x$ remove any spurious effects from the corners of a rectangular geometry. 
This effectively cylinderical system supports helical modes at the top (at $y = 10 a$) and bottom (at $y = a$) boundaries. 

Proximity induced superconductivity is conventionally introduced in this setup by adding a uniform order parameter $\Delta$ to the BdG representation of eq.~(\ref{H_BHZ})~\cite{Vivek_chiral}. 
Here, we allow the order parameter to vary independently on each site ${\bf i}$, and consider a term of the form, 
\begin{align}
  \HS = \sum_{\bf i, \sigma} \Delta_U(\bf i,\sigma) &c^{\dagger}_{{\bf i}\sigma \uparrow} c^{\dagger}_{{\bf i} \sigma \downarrow} + \text{H.c.}, 
  \, \text{ where} \\
    \Delta_U(\bf i,\sigma) &= -U \langle c_{{\bf i}\sigma\downarrow} c_{{\bf i}\sigma \uparrow} \rangle. 
    \label{Delta}   
\end{align}
Note that the order parameter is assumed to be local spatially as well as in orbital space ($\sigma$). 
While the precise orbital structure of the induced pairing would depend on details of the coupling between the superconductor and the QSH edge, we do not expect this to affect our results qualitatively~\cite{Annica2011}. 
$\HS$ may be generated through the mean-field decomposition of an attractive onsite interaction, $H_{U} = -U \sum_{\bf i, \sigma} n_{{\bf i},\sigma,\uparrow} n_{{\bf i},\sigma,\downarrow}$, in the pairing channel (here $n_{\bf i, \sigma, \tau} = c^{\dagger}_{{\bf i}\sigma \tau} c_{{\bf i}\sigma \tau}$ is the local number operator). 
Modelling the source of superconductivity as an intrinsic attraction, rather than an extrinsic superconductor allows us to approach a larger system size than would be possible otherwise. 
As described below, proximity effect in topological insulators may be modelled in this way if the induced pairing amplitude is smaller than the bulk band gap.
Here, we choose sufficiently small values of the attraction $U$ (specifically, $U < 3.5 t$) such that this condition is satisfied and there is effectively no pairing deep in the bulk, as expected for the proximitized system. 
We have also verified that in the absence of a boost ($\gp = 0$), our results qualitatively match those of Ref.~\onlinecite{Annica2011} where a more specific  model of the proximity effect was employed. 

We find the self-consistent ground state of the system through an iterative procedure with randomized initial conditions (the details were presented in Ref.~\onlinecite{Suman2020}). 
For all parameters, the calculation was repeated twice. 
First, with a spatially uniform initial order parameter, and second, when the initial $\Delta_{U}$ was allowed to assume a different value at each site. 
The magnitude (phase) of $\Delta_{U}$ was drawn from a uniform distribution in the range $[0, 0.5 t]$ ($[-\pi, \pi]$). 
In the cases where these two initial choices converged to a different state, the one with the lower energy (defined as $\langle \HB + H_{U} \rangle$) was selected as the ground state. 
For the results presented in this Letter, we used parameter values $U = 3 t$, $M_{0} = -2 t$, $B/a^2 = -t$ and varied the boost in the range $0 \leq \gp \leq 0.3 t$. 

As the QSH insulator supports low energy modes, available for pairing, only at its boundary, if the superconductivity was induced through tunnel coupling to an external superconductor, then the proximity effect would induce an effective pairing only among the edge modes and leave the bulk untouched. 
Here, we assume that the induced pairing amplitude is much smaller than the bulk band gap. 
In other words, the proximity induced order parameter only penetrates the QSH insulator upto the transverse width of the edge modes. 
This may be contrasted with the proximity effect in a normal metal, where the order parameter decays inside the metal over a distance comparable to the superconducting coherence length. 
The difference between the two cases arises because the bulk band gap (in the QSH insulator) suppresses superconductivity and does not allow the order parameter to penetrate deeper than the width of the edge modes (which is much smaller than the coherence length). 
If superconductivity is induced, in the QSH insulator, through a sufficiently small intrinsic attraction, then the pairing amplitude will be similarly suppressed in the bulk and dominantly affect only the boundary modes.  
This is qualitatively identical to the proximity effect of an extrinsic superconductor.  
Therefore, we believe that it is reasonable to model the proximity effect in QSH insulators (but not in generic metals) through an intrinsic attraction as long as the pairing amplitude of the edge is smaller than the bulk band gap, or equivalently, the superconducting coherence length is larger than the width of the edge modes.

\begin{figure}[t]
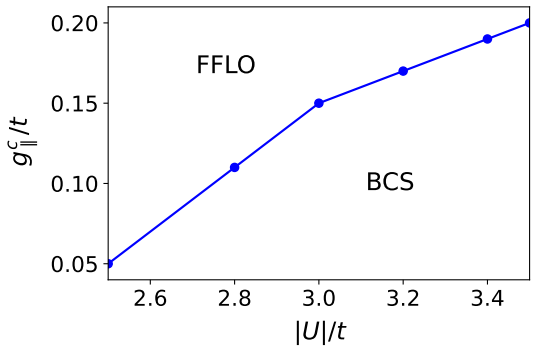

    \centering
    \includegraphics[width=0.9\columnwidth]{{{critical_g1_vs_U_for_M0_-2.0_Nx_40.pdf}}}
    \caption{Variation of the critical boost $\gp^{c}$ (at which the BCS to FFLO transition occurs) with the attraction $U$. 
    The pairing amplitude induced at the edge (in absence of the boost) is proportional to $U$. 
    The variation of $\gp^{c}$ with $U$ follows that of the pairing amplitude. 
    Note that in the range of $U$ considered here, the pairing is exponentially suppressed deep in the bulk, as expected for a proximitized system. 
    The other parameter values used here were the same as that for Fig.~\ref{fig:1}.}
    \label{fig:gpc}
\end{figure}

\section{Results}

\subsection{Nature of Induced Superconductivity}

To analyze the self-consistent ground state, we first consider the spatial variation of (the magnitude and phase of) the local pairing amplitude averaged over the orbitals, $F_U({\bf i})=\sum_{\sigma=A,B} \Delta_U({\bf i},\sigma)/2$. 
Upon performing a Fourier transform along the direction of the edge, the amplitude may be decomposed into an $s$ and $p$-wave components, which helps clarify the topological character of the induced pairing~\cite{Annica2011}. 
Here, we shall focus on the real-space behavior of $F_{U}$ which helps distinguish the BCS and FFLO phases very clearly.

Figure~\ref{fig:1} shows the change in the behavior of $F_{U}({\bf i})$ with $\gp$. 
For all values of $\gp$, the pairing amplitude decays exponentially in the bulk and remains uniform along the edge (along $x$) as expected for a proximitized setup. 
However, the variation of the phase of the order parameter along the edge is markedly different for low and high values of the boost. 
In the absence of a boost ($\gp = 0$), the induced order parameter is uniform along $x$, indicating a BCS-like superconducting state (as expected). 
For small values of the boost, we find that the induced pairing does not change qualitatively.  

Interestingly, for sufficiently large values of the boost ($\gp \gtrsim 0.15 t$ in Fig.~\ref{fig:1}), we find that the phase of the order parameter starts to vary along $x$ and that it completes an integer number of cycles from $0$ to $2\pi$ over the length of the edge. 
This winding integer is 1 for $\gp = 0.2 t$ as shown in Fig.~\ref{fig:1} and increases with $\gp$. 
This is the FFLO superconducting phase that supports finite momentum Cooper pairs~\cite{FFLO-1,FFLO-2}.  

The critical value of the boost at which this transition occurs ($\gp^{c} \sim 0.15 t$ in Fig.~\ref{fig:1}) is around the value of the pairing amplitude induced at the edge (in the absence of the boost). 
Figure~\ref{fig:1} shows results for $U = 3 t$ for which this amplitude is $\sim 0.2 t$. 
As shown in Fig.~\ref{fig:gpc}, the critical boost $\gp^{c}$ varies monotonically with the attraction $U$, which is a proxy for the induced pairing amplitude.

\begin{figure}[t]
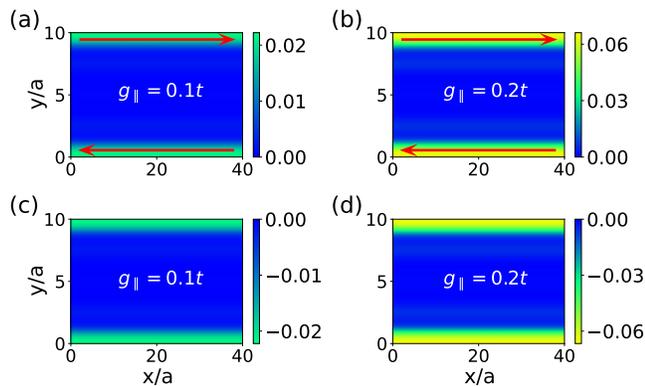

    \centering
    \includegraphics[width=1\columnwidth]{{{current_and_magnetization_dist_del_0.0.pdf}}}
    \caption{ Spatial distribution of the (a,b) current in units of $\frac{\hbar}{e}$ and (c,d) magnetization [defined in eqs.~\ref{Mag},\ref{Current}] induced by a finite boost in the absence of superconductivity. 
    Panels (a,c) correspond to $\gp = 0.1 t$, while (b,d) correspond to $\gp = 0.2 t$.  
    In both cases, the boost is much smaller than the bulk band gap, and consequently it only affects the boundaries of the system.  
    Note that the color in panels (a,b) depicts the magnitude of the current, while the arrows mark its direction. 
    In (c,d) the color depicts the magnetization (along with its sign). 
    While the magnetization has the same sign on the two edges, the direction of the persistent current that flows along the edge is opposite (as depicted by the arrows). 
    The parameter values used here were same as that for Fig.~\ref{fig:1}.} 
    \label{fig:noSC}
\end{figure}

\subsection{Persistent Current and Magnetization}

In the absence of superconductivity, the application of a boost leads to a charge imbalance between the two spin sectors, which manifests as a finite magnetization along the direction of spin quantization of the helical edge. 
Furthermore, due to the spin-momentum locking at the helical edge, the boost also leads to a finite charge current flowing along the edge. 
In order to understand the fate of these properties in presence of induced superconductivity, we evaluated the local magnetization ($\langle M({\bf i}) \rangle$ at each site ${\bf i}$) and the bond current ($\langle J_{{\bf i j}} \rangle$ from site ${\bf i}$ to ${\bf j}$) in the self-consistent ground state. 
These observables are given by, 
\begin{align}
  M({\bf i}) &= \sum_{\sigma} \Big[ n_{{\bf i} \sigma \uparrow} -  n_{{\bf i} \sigma \downarrow} \Big], \label{Mag} \\ 
  J_{{\bf ij}} &= \frac{2e}{\hbar} \sum_{\sigma,\tau,\sigma',\tau'} \text{Im} \Big[t_{{\bf ji}}(\sigma,\tau;\sigma',\tau') c^{\dagger}_{{\bf j} \sigma \tau} c_{{\bf i} \sigma' \tau'} \Big]. \label{Current}
\end{align}
Here, $n_{\bf i, \sigma, \tau} = c^{\dagger}_{{\bf i}\sigma \tau} c_{{\bf i}\sigma \tau}$ is the local fermion number, and $t_{{\bf ji}}(\sigma,\tau;\sigma',\tau')$ is the hopping between $|\sigma, \tau\rangle$ at site ${\bf i}$ and $|\sigma^{\prime}, \tau^{\prime} \rangle$ at site ${\bf j}$ in the discretized version of (\ref{H_BHZ}). 

Figure~\ref{fig:noSC} depicts the magnetization and current induced by a boost (for $\gp = 0.1 t$ and $0.2t$) applied uniformly throughout the system in the absence of superconductivity. 
A uniform boost shifts the entire 2D band structure in energy. 
For $\gp \ll 2 |M_{0}| = 4 t$, only the edge modes cross the chemical potential (pinned to zero throughout the calculation) and hence the effects of charge imbalance (between the spin sectors) are restricted to the boundaries, while the bulk remains inert. 
With our sign convention, a positive $\gp$ increases the charge in the spin-down sector, giving rise to negative magnetization [Fig.~\ref{fig:noSC}(c)] at the boundaries. 
The opposite helicities of the top and bottom edges lead to a persistent current in the opposite directions along the two boundaries. 
Note that the magnitude of the current, depicted by the color in Fig.~\ref{fig:noSC}(a,b), depends on the applied boost. 

A notable feature of topological band structures is that a sharp or smooth domain wall in the chemical potential leads to additional subgap modes which carry current along these interfaces~\cite{Pachos_2020,Sun_2021}. 
This is the case even if the change in chemical potential is much smaller than the bulk band gap. 
Consequently, applying the boost in a finite region which includes the two boundaries generates additional interface currents deep in the bulk. 
To simplify our analysis, here we ignore the interplay of these additional states with induced superconductivity, and apply the boost uniformly throughout the setup. 
Additionally, from the point of view of experimental feasibility, realizing a uniform boost is probably simpler than applying a Zeeman field just at the edge.

\begin{figure}[t]
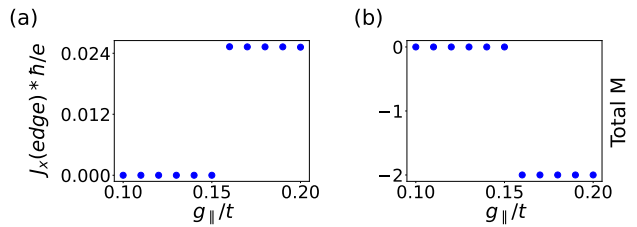

    \centering
    \includegraphics[width=1\columnwidth]{{{current_and_magnetization_vs_g1_V_3.0.pdf}}}
    \caption{Variation of the (a) current at the top edge ($y = 10 a$) and (b) total magnetization of the system with the boost in the presence of induced pairing. 
    Here, we used $U = 3 t$ which leads to $\gp^{c} \sim 0.15 t$.
    The sharp jump in the current and magnetization appears for $\gp \sim \gp^{c}$ at which the FFLO phase emerges. 
    Notably, the FFLO ground state carries a finite current as well as magnetization. 
    The parameter values used here were same as that for Fig~\ref{fig:1}.} 
    \label{fig:Cur_Mag_SC}
\end{figure}

Figure~\ref{fig:Cur_Mag_SC} depicts the variation of the total magnetization ($\sum_{\bf i} M(\bf i)$, $M(\bf i)$ is defined in eq.\ref{Mag}) and the bond current at one of the boundaries with boost in presence of induced superconductivity ($U = 3t$ with $\gp^{c} \sim 0.15 t$). 
Evidently, the BCS to FFLO phase transition is accompanied by a sharp jump in the current and magnetization as well. 
Interestingly, the BCS-like ground state does not support any ground state current and magnetization even at finite boost. 
This suggests that the self-consistent calculation modifies the boosted band structure such that the charge-imbalance generated by the boost is removed. 
In contrast, the FFLO ground state does support finite magnetization and current along the edge, implying that the boost induced charge-imbalance is not completely removed. 
The magnitude of the current in FFLO phase is smaller than what was induced by the boost in absence of superconductivity. 
For instance, at $\gp = 0.2t$ the current is $\sim 0.06$ $\frac{\hbar}{e}$ without pairing [Fig.~\ref{fig:noSC}(b)] and reduces to $\sim 0.02$ $\frac{\hbar}{e}$ in the FFLO phase [Fig.~\ref{fig:Cur_Mag_SC}(a)]. 
Hence, our results indicate that the FFLO phase becomes energetically favorable at large $\gp$, because the system is able to partially remove the persistent current induced by the boost through the supercurrent arising from the spatial variation (of the phase) of the order parameter. 

\begin{figure}[t]
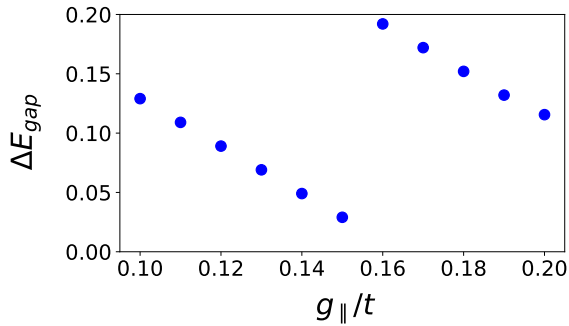

    \centering
    \includegraphics[width=0.9\columnwidth]{{{Energy_gap_vs_g1_V_3.0.pdf}}}
    \caption{Variation of the single-particle BdG gap with the boost.  
    For small values of the boost, the ground state is in a BCS-like phase which is adiabatically connected to the state at $\gp = 0$. 
    The sharp jump in the single-particle gap (at $\gp = \gp^{c} \sim 0.15 t$) occurs at the first-order transition to the FFLO phase. 
    The gap appears to close within the BCS phase, even though the induced pairing amplitude remains almost constant. 
    This suggests that the boost tends to induce an indirect gap closing in the BdG spectrum, and that the FFLO phase emerges around the gap closing point. 
    The parameter values used here were same as that for Fig~\ref{fig:1}.} 
    \label{fig:gap}
\end{figure}

\section{Discussion and Conclusions}

It is well understood that, in the absence of a boost, an effective 1D Hamiltonian with uniform pairing is sufficient to capture the basic phenomenology of proximity induced superconductivity at the helical edge~\cite{Annica2011}. 
This is because a self-consistent calculation `only' leads to minor renormalization of the model parameters. 
In  sharp contrast, as described below, our results demonstrate the importance of a self-consistent calculation and retaining the 2D bulk band structure in the presence of a boost. 

Earlier studies~\cite{Vivek_chiral,Meyer2015}, which employed a 1D model of the boosted helical edge with a (non-self-consistent) uniform order parameter ($\Delta$), found that the boost induces an indirect gap closing in the BdG spectrum at $\gp = \Delta$. 
For $\gp > \Delta$, these studies predicted the emergence of an intriguing gapless superconducting phase. 
Figure~\ref{fig:gap} shows the self-consistent single-particle BdG gap of our setup as a function of the boost. 
Evidently, a sharp jump occurs in the variation of the gap with the boost at $\gp = \gp^{c}$ (the BCS to FFLO transition point).  
Within the BCS phase the gap clearly decreases monotonically with $\gp$, and appears to close around $\gp^{c}$. 
Notably, the induced pairing amplitude does not change significantly with the boost within the BCS phase. 
Hence, our self-consistent results are seemingly in agreement with the picture of an indirect gap closing, and a uniform order parameter, as suggested by the previous works. 
However, our analysis also demonstrates that the predicted gapless superconducting phase is not stable, and that it gives way to an FFLO phase instead. 
We emphasize here, that a non-self-consistent calculation cannot capture the emergence of an FFLO phase. 

As pointed out earlier, in addition to creating a charge-imbalance, a finite boost also shifts the Dirac cone of the helical edge state to finite momentum. 
The persistent current generated by the boost may also be attributed to this shift of the Dirac cone.  
An earlier self-consistent calculation within an effective 1D model of the helical edge (which does capture the shift of the cone) found that the FFLO phase may emerge even for infinitesimal boost~\cite{Meyer2015}, and interestingly does not carry any net current. 
The FFLO supercurrent cancels the boost-induced persistent current perfectly in these analyses. 
The differences between these results and ours (as presented in Figs.~\ref{fig:gpc} and \ref{fig:Cur_Mag_SC}) point to the importance of retaining the complete 2D band structure in the calculation. 
By including all the states in the calculation, we are able to fully account for the charge-imbalance induced by the boost. 
This imbalance does not appear in 1D models because (typically) the linear edge spectrum is terminated at some arbitrary UV cutoff, and the shift of this cutoff (induced by the boost) is ignored.
As our analysis shows, the current and magnetization induced by this charge imbalance affect the energies of different forms of pairing, and this leads to the emergence of a BCS-like (FFLO-like) phase at low (high) values of the boost.

Topological superconductivity has been studied extensively in the context of 1D Fermionic systems with strong spin-orbit coupling. 
A previous self-consistent analysis found that an FFLO phase may be induced in these setups through the application of a Zeeman field, but only if the field is larger than a finite cutoff~\cite{Chunlei_2014}. 
A subsequent DMRG calculation revealed that this result does not change qualitatively even when fluctuation effects beyond mean-field theory are accounted for~\cite{Monalisa_2021}. 
It is interesting to note the qualitative similarity of these results with ours, despite the differences between the helical edge states and the 1D Fermi gases. 
Notably, these works employed a 1D lattice model which, similar to our work, provides access to the full band structure and accounts for the charge imbalance generated by the Zeeman term.
Finite momentum pairing states (similar to FFLO) may also be induced at the edges of a quantum anomalous Hall insulator via the proximity effect~\cite{Tanaka_QAH_JJ}.

In conclusion, we studied the nature of the superconductivity induced in proximitized helical edges of a quantum spin Hall insulator in the presence of a boost. 
Our self-consistent analysis accounts fully for the charge-imbalance induced by the boost and its effect of the induced pairing. 
We find that a BCS-like phase, with uniform order parameter, persists upto a finite boost. 
If the applied boost is stronger than this cutoff, then an FFLO phase emerges, which supports a finite magnetization and current at the edge. 
The latter are signatures of a leftover charge-imbalance which was induced by the boost, but not fully removed by the self-consistent pairing. 
Our work not only motivates a future detailed analysis of the topological character of the induced pairing in presence of a boost, but also inspires a revision of the procedures employed in analyses of effective 1D models.  
Additionally, it may be interesting to consider the possibility of induced odd-frequency pairing~\cite{Tanaka_OddFrequency,Tanaka_Review}, as well as the interplay of the inverse-proximity effect~\cite{MoonJip_2017} and the induced FFLO phase at the 1D helical edge. 

We acknowledge Vivekananda Adak and Aabir Mukhopadhyay for useful scientific discussions. SJD wants to acknowledge ICTS for its hospitality, funding, and kind support toward academic collaboration. 
U.K. was supported by a fellowship from the Israel Science Foundation (ISF, Grant No. 993/19). 

\bibliography{majorana}

\begin{thebibliography}{40}%
\makeatletter
\providecommand \@ifxundefined [1]{%
 \@ifx{#1\undefined}
}%
\providecommand \@ifnum [1]{%
 \ifnum #1\expandafter \@firstoftwo
 \else \expandafter \@secondoftwo
 \fi
}%
\providecommand \@ifx [1]{%
 \ifx #1\expandafter \@firstoftwo
 \else \expandafter \@secondoftwo
 \fi
}%
\providecommand \natexlab [1]{#1}%
\providecommand \enquote  [1]{``#1''}%
\providecommand \bibnamefont  [1]{#1}%
\providecommand \bibfnamefont [1]{#1}%
\providecommand \citenamefont [1]{#1}%
\providecommand \href@noop [0]{\@secondoftwo}%
\providecommand \href [0]{\begingroup \@sanitize@url \@href}%
\providecommand \@href[1]{\@@startlink{#1}\@@href}%
\providecommand \@@href[1]{\endgroup#1\@@endlink}%
\providecommand \@sanitize@url [0]{\catcode `\\12\catcode `\$12\catcode
  `\&12\catcode `\#12\catcode `\^12\catcode `\_12\catcode `\%12\relax}%
\providecommand \@@startlink[1]{}%
\providecommand \@@endlink[0]{}%
\providecommand \url  [0]{\begingroup\@sanitize@url \@url }%
\providecommand \@url [1]{\endgroup\@href {#1}{\urlprefix }}%
\providecommand \urlprefix  [0]{URL }%
\providecommand \Eprint [0]{\href }%
\providecommand \doibase [0]{https://doi.org/}%
\providecommand \selectlanguage [0]{\@gobble}%
\providecommand \bibinfo  [0]{\@secondoftwo}%
\providecommand \bibfield  [0]{\@secondoftwo}%
\providecommand \translation [1]{[#1]}%
\providecommand \BibitemOpen [0]{}%
\providecommand \bibitemStop [0]{}%
\providecommand \bibitemNoStop [0]{.\EOS\space}%
\providecommand \EOS [0]{\spacefactor3000\relax}%
\providecommand \BibitemShut  [1]{\csname bibitem#1\endcsname}%
\let\auto@bib@innerbib\@empty
\bibitem [{\citenamefont {Kitaev}(2001)}]{Kitaev_2001}%
  \BibitemOpen
  \bibfield  {author} {\bibinfo {author} {\bibfnamefont {A.~Y.}\ \bibnamefont
  {Kitaev}},\ }\bibfield  {title} {\bibinfo {title} {Unpaired majorana fermions
  in quantum wires},\ }\href {https://doi.org/10.1070/1063-7869/44/10S/S29}
  {\bibfield  {journal} {\bibinfo  {journal} {Physics-Uspekhi}\ }\textbf
  {\bibinfo {volume} {44}},\ \bibinfo {pages} {131} (\bibinfo {year}
  {2001})}\BibitemShut {NoStop}%
\bibitem [{\citenamefont {Qi}\ and\ \citenamefont
  {Zhang}(2011)}]{Zhang_2011_rev}%
  \BibitemOpen
  \bibfield  {author} {\bibinfo {author} {\bibfnamefont {X.-L.}\ \bibnamefont
  {Qi}}\ and\ \bibinfo {author} {\bibfnamefont {S.-C.}\ \bibnamefont {Zhang}},\
  }\bibfield  {title} {\bibinfo {title} {Topological insulators and
  superconductors},\ }\href {https://doi.org/10.1103/RevModPhys.83.1057}
  {\bibfield  {journal} {\bibinfo  {journal} {Rev. Mod. Phys.}\ }\textbf
  {\bibinfo {volume} {83}},\ \bibinfo {pages} {1057} (\bibinfo {year}
  {2011})}\BibitemShut {NoStop}%
\bibitem [{\citenamefont {Beenakker}(2013)}]{Beenakker2013_rev}%
  \BibitemOpen
  \bibfield  {author} {\bibinfo {author} {\bibfnamefont {C.}~\bibnamefont
  {Beenakker}},\ }\bibfield  {title} {\bibinfo {title} {Search for majorana
  fermions in superconductors},\ }\href
  {https://doi.org/10.1146/annurev-conmatphys-030212-184337} {\bibfield
  {journal} {\bibinfo  {journal} {Annual Review of Condensed Matter Physics}\
  }\textbf {\bibinfo {volume} {4}},\ \bibinfo {pages} {113} (\bibinfo {year}
  {2013})}\BibitemShut {NoStop}%
\bibitem [{\citenamefont {Kitaev}(2003)}]{Kitaev_2003}%
  \BibitemOpen
  \bibfield  {author} {\bibinfo {author} {\bibfnamefont {A.}~\bibnamefont
  {Kitaev}},\ }\bibfield  {title} {\bibinfo {title} {Fault-tolerant quantum
  computation by anyons},\ }\href
  {https://doi.org/https://doi.org/10.1016/S0003-4916(02)00018-0} {\bibfield
  {journal} {\bibinfo  {journal} {Annals of Physics}\ }\textbf {\bibinfo
  {volume} {303}},\ \bibinfo {pages} {2} (\bibinfo {year} {2003})}\BibitemShut
  {NoStop}%
\bibitem [{\citenamefont {Nayak}\ \emph {et~al.}(2008)\citenamefont {Nayak},
  \citenamefont {Simon}, \citenamefont {Stern}, \citenamefont {Freedman},\ and\
  \citenamefont {Das~Sarma}}]{Nayak_RMP_2008}%
  \BibitemOpen
  \bibfield  {author} {\bibinfo {author} {\bibfnamefont {C.}~\bibnamefont
  {Nayak}}, \bibinfo {author} {\bibfnamefont {S.~H.}\ \bibnamefont {Simon}},
  \bibinfo {author} {\bibfnamefont {A.}~\bibnamefont {Stern}}, \bibinfo
  {author} {\bibfnamefont {M.}~\bibnamefont {Freedman}},\ and\ \bibinfo
  {author} {\bibfnamefont {S.}~\bibnamefont {Das~Sarma}},\ }\bibfield  {title}
  {\bibinfo {title} {Non-abelian anyons and topological quantum computation},\
  }\href {https://doi.org/10.1103/RevModPhys.80.1083} {\bibfield  {journal}
  {\bibinfo  {journal} {Rev. Mod. Phys.}\ }\textbf {\bibinfo {volume} {80}},\
  \bibinfo {pages} {1083} (\bibinfo {year} {2008})}\BibitemShut {NoStop}%
\bibitem [{\citenamefont {Sau}\ \emph {et~al.}(2010)\citenamefont {Sau},
  \citenamefont {Lutchyn}, \citenamefont {Tewari},\ and\ \citenamefont
  {Das~Sarma}}]{Sau2010}%
  \BibitemOpen
  \bibfield  {author} {\bibinfo {author} {\bibfnamefont {J.~D.}\ \bibnamefont
  {Sau}}, \bibinfo {author} {\bibfnamefont {R.~M.}\ \bibnamefont {Lutchyn}},
  \bibinfo {author} {\bibfnamefont {S.}~\bibnamefont {Tewari}},\ and\ \bibinfo
  {author} {\bibfnamefont {S.}~\bibnamefont {Das~Sarma}},\ }\bibfield  {title}
  {\bibinfo {title} {Generic new platform for topological quantum computation
  using semiconductor heterostructures},\ }\href
  {https://doi.org/10.1103/PhysRevLett.104.040502} {\bibfield  {journal}
  {\bibinfo  {journal} {Phys. Rev. Lett.}\ }\textbf {\bibinfo {volume} {104}},\
  \bibinfo {pages} {040502} (\bibinfo {year} {2010})}\BibitemShut {NoStop}%
\bibitem [{\citenamefont {Oreg}\ \emph {et~al.}(2010)\citenamefont {Oreg},
  \citenamefont {Refael},\ and\ \citenamefont {von Oppen}}]{Oreg2010}%
  \BibitemOpen
  \bibfield  {author} {\bibinfo {author} {\bibfnamefont {Y.}~\bibnamefont
  {Oreg}}, \bibinfo {author} {\bibfnamefont {G.}~\bibnamefont {Refael}},\ and\
  \bibinfo {author} {\bibfnamefont {F.}~\bibnamefont {von Oppen}},\ }\bibfield
  {title} {\bibinfo {title} {Helical liquids and majorana bound states in
  quantum wires},\ }\href {https://doi.org/10.1103/PhysRevLett.105.177002}
  {\bibfield  {journal} {\bibinfo  {journal} {Phys. Rev. Lett.}\ }\textbf
  {\bibinfo {volume} {105}},\ \bibinfo {pages} {177002} (\bibinfo {year}
  {2010})}\BibitemShut {NoStop}%
\bibitem [{\citenamefont {Sato}\ and\ \citenamefont
  {Ando}(2017)}]{Ando_Review_2017}%
  \BibitemOpen
  \bibfield  {author} {\bibinfo {author} {\bibfnamefont {M.}~\bibnamefont
  {Sato}}\ and\ \bibinfo {author} {\bibfnamefont {Y.}~\bibnamefont {Ando}},\
  }\bibfield  {title} {\bibinfo {title} {Topological superconductors: a
  review},\ }\href {https://doi.org/10.1088/1361-6633/aa6ac7} {\bibfield
  {journal} {\bibinfo  {journal} {Rep. Prog. Phys.}\ }\textbf {\bibinfo
  {volume} {80}},\ \bibinfo {pages} {076501} (\bibinfo {year}
  {2017})}\BibitemShut {NoStop}%
\bibitem [{\citenamefont {Flensberg}\ \emph {et~al.}(2021)\citenamefont
  {Flensberg}, \citenamefont {von Oppen},\ and\ \citenamefont
  {Stern}}]{Stern_Review_2021}%
  \BibitemOpen
  \bibfield  {author} {\bibinfo {author} {\bibfnamefont {K.}~\bibnamefont
  {Flensberg}}, \bibinfo {author} {\bibfnamefont {F.}~\bibnamefont {von
  Oppen}},\ and\ \bibinfo {author} {\bibfnamefont {A.}~\bibnamefont {Stern}},\
  }\bibfield  {title} {\bibinfo {title} {Engineered platforms for topological
  superconductivity and majorana zero modes},\ }\href
  {https://doi.org/10.1038/s41578-021-00336-6} {\bibfield  {journal} {\bibinfo
  {journal} {Nature Reviews Materials}\ }\textbf {\bibinfo {volume} {6}},\
  \bibinfo {pages} {944} (\bibinfo {year} {2021})}\BibitemShut {NoStop}%
\bibitem [{\citenamefont {M\ae{}land}\ and\ \citenamefont
  {Sudb\o{}}(2023)}]{Kristian2023}%
  \BibitemOpen
  \bibfield  {author} {\bibinfo {author} {\bibfnamefont {K.}~\bibnamefont
  {M\ae{}land}}\ and\ \bibinfo {author} {\bibfnamefont {A.}~\bibnamefont
  {Sudb\o{}}},\ }\bibfield  {title} {\bibinfo {title} {Topological
  superconductivity mediated by skyrmionic magnons},\ }\href
  {https://doi.org/10.1103/PhysRevLett.130.156002} {\bibfield  {journal}
  {\bibinfo  {journal} {Phys. Rev. Lett.}\ }\textbf {\bibinfo {volume} {130}},\
  \bibinfo {pages} {156002} (\bibinfo {year} {2023})}\BibitemShut {NoStop}%
\bibitem [{\citenamefont {Wu}\ \emph {et~al.}(2006)\citenamefont {Wu},
  \citenamefont {Bernevig},\ and\ \citenamefont {Zhang}}]{Wu2006}%
  \BibitemOpen
  \bibfield  {author} {\bibinfo {author} {\bibfnamefont {C.}~\bibnamefont
  {Wu}}, \bibinfo {author} {\bibfnamefont {B.~A.}\ \bibnamefont {Bernevig}},\
  and\ \bibinfo {author} {\bibfnamefont {S.-C.}\ \bibnamefont {Zhang}},\
  }\bibfield  {title} {\bibinfo {title} {Helical liquid and the edge of quantum
  spin hall systems},\ }\href {https://doi.org/10.1103/PhysRevLett.96.106401}
  {\bibfield  {journal} {\bibinfo  {journal} {Phys. Rev. Lett.}\ }\textbf
  {\bibinfo {volume} {96}},\ \bibinfo {pages} {106401} (\bibinfo {year}
  {2006})}\BibitemShut {NoStop}%
\bibitem [{\citenamefont {Xu}\ and\ \citenamefont {Moore}(2006)}]{Xu2006}%
  \BibitemOpen
  \bibfield  {author} {\bibinfo {author} {\bibfnamefont {C.}~\bibnamefont
  {Xu}}\ and\ \bibinfo {author} {\bibfnamefont {J.~E.}\ \bibnamefont {Moore}},\
  }\bibfield  {title} {\bibinfo {title} {Stability of the quantum spin hall
  effect: Effects of interactions, disorder, and ${\mathbb{z}}_{2}$ topology},\
  }\href {https://doi.org/10.1103/PhysRevB.73.045322} {\bibfield  {journal}
  {\bibinfo  {journal} {Phys. Rev. B}\ }\textbf {\bibinfo {volume} {73}},\
  \bibinfo {pages} {045322} (\bibinfo {year} {2006})}\BibitemShut {NoStop}%
\bibitem [{\citenamefont {Kane}\ and\ \citenamefont
  {Mele}(2005{\natexlab{a}})}]{Kane2005-1}%
  \BibitemOpen
  \bibfield  {author} {\bibinfo {author} {\bibfnamefont {C.~L.}\ \bibnamefont
  {Kane}}\ and\ \bibinfo {author} {\bibfnamefont {E.~J.}\ \bibnamefont
  {Mele}},\ }\bibfield  {title} {\bibinfo {title} {Quantum spin hall effect in
  graphene},\ }\href {https://doi.org/10.1103/PhysRevLett.95.226801} {\bibfield
   {journal} {\bibinfo  {journal} {Phys. Rev. Lett.}\ }\textbf {\bibinfo
  {volume} {95}},\ \bibinfo {pages} {226801} (\bibinfo {year}
  {2005}{\natexlab{a}})}\BibitemShut {NoStop}%
\bibitem [{\citenamefont {Kane}\ and\ \citenamefont
  {Mele}(2005{\natexlab{b}})}]{Kane2005-2}%
  \BibitemOpen
  \bibfield  {author} {\bibinfo {author} {\bibfnamefont {C.~L.}\ \bibnamefont
  {Kane}}\ and\ \bibinfo {author} {\bibfnamefont {E.~J.}\ \bibnamefont
  {Mele}},\ }\bibfield  {title} {\bibinfo {title} {${Z}_{2}$ topological order
  and the quantum spin hall effect},\ }\href
  {https://doi.org/10.1103/PhysRevLett.95.146802} {\bibfield  {journal}
  {\bibinfo  {journal} {Phys. Rev. Lett.}\ }\textbf {\bibinfo {volume} {95}},\
  \bibinfo {pages} {146802} (\bibinfo {year} {2005}{\natexlab{b}})}\BibitemShut
  {NoStop}%
\bibitem [{\citenamefont {Bernevig}\ \emph {et~al.}(2006)\citenamefont
  {Bernevig}, \citenamefont {Hughes},\ and\ \citenamefont
  {Zhang}}]{Bernevig2006}%
  \BibitemOpen
  \bibfield  {author} {\bibinfo {author} {\bibfnamefont {B.~A.}\ \bibnamefont
  {Bernevig}}, \bibinfo {author} {\bibfnamefont {T.~L.}\ \bibnamefont
  {Hughes}},\ and\ \bibinfo {author} {\bibfnamefont {S.-C.}\ \bibnamefont
  {Zhang}},\ }\bibfield  {title} {\bibinfo {title} {Quantum spin hall effect
  and topological phase transition in hgte quantum wells},\ }\href
  {https://doi.org/10.1126/science.1133734} {\bibfield  {journal} {\bibinfo
  {journal} {Science}\ }\textbf {\bibinfo {volume} {314}},\ \bibinfo {pages}
  {1757} (\bibinfo {year} {2006})}\BibitemShut {NoStop}%
\bibitem [{\citenamefont {K{\"o}nig}\ \emph {et~al.}(2007)\citenamefont
  {K{\"o}nig}, \citenamefont {Wiedmann}, \citenamefont {Br{\"u}ne},
  \citenamefont {Roth}, \citenamefont {Buhmann}, \citenamefont {Molenkamp},
  \citenamefont {Qi},\ and\ \citenamefont {Zhang}}]{Konig2007}%
  \BibitemOpen
  \bibfield  {author} {\bibinfo {author} {\bibfnamefont {M.}~\bibnamefont
  {K{\"o}nig}}, \bibinfo {author} {\bibfnamefont {S.}~\bibnamefont {Wiedmann}},
  \bibinfo {author} {\bibfnamefont {C.}~\bibnamefont {Br{\"u}ne}}, \bibinfo
  {author} {\bibfnamefont {A.}~\bibnamefont {Roth}}, \bibinfo {author}
  {\bibfnamefont {H.}~\bibnamefont {Buhmann}}, \bibinfo {author} {\bibfnamefont
  {L.~W.}\ \bibnamefont {Molenkamp}}, \bibinfo {author} {\bibfnamefont {X.-L.}\
  \bibnamefont {Qi}},\ and\ \bibinfo {author} {\bibfnamefont {S.-C.}\
  \bibnamefont {Zhang}},\ }\bibfield  {title} {\bibinfo {title} {Quantum spin
  hall insulator state in hgte quantum wells},\ }\href
  {https://doi.org/10.1126/science.1148047} {\bibfield  {journal} {\bibinfo
  {journal} {Science}\ }\textbf {\bibinfo {volume} {318}},\ \bibinfo {pages}
  {766} (\bibinfo {year} {2007})}\BibitemShut {NoStop}%
\bibitem [{\citenamefont {Liu}\ \emph {et~al.}(2008)\citenamefont {Liu},
  \citenamefont {Hughes}, \citenamefont {Qi}, \citenamefont {Wang},\ and\
  \citenamefont {Zhang}}]{Liu2008}%
  \BibitemOpen
  \bibfield  {author} {\bibinfo {author} {\bibfnamefont {C.}~\bibnamefont
  {Liu}}, \bibinfo {author} {\bibfnamefont {T.~L.}\ \bibnamefont {Hughes}},
  \bibinfo {author} {\bibfnamefont {X.-L.}\ \bibnamefont {Qi}}, \bibinfo
  {author} {\bibfnamefont {K.}~\bibnamefont {Wang}},\ and\ \bibinfo {author}
  {\bibfnamefont {S.-C.}\ \bibnamefont {Zhang}},\ }\bibfield  {title} {\bibinfo
  {title} {Quantum spin hall effect in inverted type-ii semiconductors},\
  }\href {https://doi.org/10.1103/PhysRevLett.100.236601} {\bibfield  {journal}
  {\bibinfo  {journal} {Phys. Rev. Lett.}\ }\textbf {\bibinfo {volume} {100}},\
  \bibinfo {pages} {236601} (\bibinfo {year} {2008})}\BibitemShut {NoStop}%
\bibitem [{\citenamefont {Fu}\ and\ \citenamefont {Kane}(2008)}]{FuKane2008}%
  \BibitemOpen
  \bibfield  {author} {\bibinfo {author} {\bibfnamefont {L.}~\bibnamefont
  {Fu}}\ and\ \bibinfo {author} {\bibfnamefont {C.~L.}\ \bibnamefont {Kane}},\
  }\bibfield  {title} {\bibinfo {title} {Superconducting proximity effect and
  majorana fermions at the surface of a topological insulator},\ }\href
  {https://doi.org/10.1103/PhysRevLett.100.096407} {\bibfield  {journal}
  {\bibinfo  {journal} {Phys. Rev. Lett.}\ }\textbf {\bibinfo {volume} {100}},\
  \bibinfo {pages} {096407} (\bibinfo {year} {2008})}\BibitemShut {NoStop}%
\bibitem [{\citenamefont {Fu}\ and\ \citenamefont
  {Kane}(2009)}]{FuKanePRB2009}%
  \BibitemOpen
  \bibfield  {author} {\bibinfo {author} {\bibfnamefont {L.}~\bibnamefont
  {Fu}}\ and\ \bibinfo {author} {\bibfnamefont {C.~L.}\ \bibnamefont {Kane}},\
  }\bibfield  {title} {\bibinfo {title} {Josephson current and noise at a
  superconductor/quantum-spin-hall-insulator/superconductor junction},\ }\href
  {https://doi.org/10.1103/PhysRevB.79.161408} {\bibfield  {journal} {\bibinfo
  {journal} {Phys. Rev. B}\ }\textbf {\bibinfo {volume} {79}},\ \bibinfo
  {pages} {161408} (\bibinfo {year} {2009})}\BibitemShut {NoStop}%
\bibitem [{\citenamefont {Adak}\ \emph {et~al.}(2022)\citenamefont {Adak},
  \citenamefont {Mukhopadhyay}, \citenamefont {De}, \citenamefont {Khanna},
  \citenamefont {Rao},\ and\ \citenamefont {Das}}]{Vivek_chiral}%
  \BibitemOpen
  \bibfield  {author} {\bibinfo {author} {\bibfnamefont {V.}~\bibnamefont
  {Adak}}, \bibinfo {author} {\bibfnamefont {A.}~\bibnamefont {Mukhopadhyay}},
  \bibinfo {author} {\bibfnamefont {S.~J.}\ \bibnamefont {De}}, \bibinfo
  {author} {\bibfnamefont {U.}~\bibnamefont {Khanna}}, \bibinfo {author}
  {\bibfnamefont {S.}~\bibnamefont {Rao}},\ and\ \bibinfo {author}
  {\bibfnamefont {S.}~\bibnamefont {Das}},\ }\bibfield  {title} {\bibinfo
  {title} {Chiral detection of majorana bound states at the edge of a quantum
  spin hall insulator},\ }\href {https://doi.org/10.1103/PhysRevB.106.045422}
  {\bibfield  {journal} {\bibinfo  {journal} {Phys. Rev. B}\ }\textbf {\bibinfo
  {volume} {106}},\ \bibinfo {pages} {045422} (\bibinfo {year}
  {2022})}\BibitemShut {NoStop}%
\bibitem [{\citenamefont {Li}\ \emph {et~al.}(2013)\citenamefont {Li},
  \citenamefont {Sheng}, \citenamefont {Shen}, \citenamefont {Shao},
  \citenamefont {Wang}, \citenamefont {Sheng},\ and\ \citenamefont
  {Xing}}]{Li2013}%
  \BibitemOpen
  \bibfield  {author} {\bibinfo {author} {\bibfnamefont {H.}~\bibnamefont
  {Li}}, \bibinfo {author} {\bibfnamefont {L.}~\bibnamefont {Sheng}}, \bibinfo
  {author} {\bibfnamefont {R.}~\bibnamefont {Shen}}, \bibinfo {author}
  {\bibfnamefont {L.~B.}\ \bibnamefont {Shao}}, \bibinfo {author}
  {\bibfnamefont {B.}~\bibnamefont {Wang}}, \bibinfo {author} {\bibfnamefont
  {D.~N.}\ \bibnamefont {Sheng}},\ and\ \bibinfo {author} {\bibfnamefont
  {D.~Y.}\ \bibnamefont {Xing}},\ }\bibfield  {title} {\bibinfo {title}
  {Stabilization of the quantum spin hall effect by designed removal of
  time-reversal symmetry of edge states},\ }\href
  {https://doi.org/10.1103/PhysRevLett.110.266802} {\bibfield  {journal}
  {\bibinfo  {journal} {Phys. Rev. Lett.}\ }\textbf {\bibinfo {volume} {110}},\
  \bibinfo {pages} {266802} (\bibinfo {year} {2013})}\BibitemShut {NoStop}%
\bibitem [{\citenamefont {Yu}\ \emph {et~al.}(2021)\citenamefont {Yu},
  \citenamefont {Chen}, \citenamefont {Gomanko}, \citenamefont {Badawy},
  \citenamefont {Bakkers}, \citenamefont {Zuo}, \citenamefont {Mourik},\ and\
  \citenamefont {Frolov}}]{Frolov2021}%
  \BibitemOpen
  \bibfield  {author} {\bibinfo {author} {\bibfnamefont {P.}~\bibnamefont
  {Yu}}, \bibinfo {author} {\bibfnamefont {J.}~\bibnamefont {Chen}}, \bibinfo
  {author} {\bibfnamefont {M.}~\bibnamefont {Gomanko}}, \bibinfo {author}
  {\bibfnamefont {G.}~\bibnamefont {Badawy}}, \bibinfo {author} {\bibfnamefont
  {E.~P. A.~M.}\ \bibnamefont {Bakkers}}, \bibinfo {author} {\bibfnamefont
  {K.}~\bibnamefont {Zuo}}, \bibinfo {author} {\bibfnamefont {V.}~\bibnamefont
  {Mourik}},\ and\ \bibinfo {author} {\bibfnamefont {S.~M.}\ \bibnamefont
  {Frolov}},\ }\bibfield  {title} {\bibinfo {title} {Non-majorana states yield
  nearly quantized conductance in proximatized nanowires},\ }\href
  {https://doi.org/10.1038/s41567-020-01107-w} {\bibfield  {journal} {\bibinfo
  {journal} {Nature Physics}\ }\textbf {\bibinfo {volume} {17}},\ \bibinfo
  {pages} {482} (\bibinfo {year} {2021})}\BibitemShut {NoStop}%
\bibitem [{\citenamefont {Das~Sarma}\ and\ \citenamefont
  {Pan}(2021)}]{DasSarma2021A}%
  \BibitemOpen
  \bibfield  {author} {\bibinfo {author} {\bibfnamefont {S.}~\bibnamefont
  {Das~Sarma}}\ and\ \bibinfo {author} {\bibfnamefont {H.}~\bibnamefont
  {Pan}},\ }\bibfield  {title} {\bibinfo {title} {Disorder-induced zero-bias
  peaks in majorana nanowires},\ }\href
  {https://doi.org/10.1103/PhysRevB.103.195158} {\bibfield  {journal} {\bibinfo
   {journal} {Phys. Rev. B}\ }\textbf {\bibinfo {volume} {103}},\ \bibinfo
  {pages} {195158} (\bibinfo {year} {2021})}\BibitemShut {NoStop}%
\bibitem [{\citenamefont {Pan}\ and\ \citenamefont
  {Das~Sarma}(2021)}]{DasSarma2021B}%
  \BibitemOpen
  \bibfield  {author} {\bibinfo {author} {\bibfnamefont {H.}~\bibnamefont
  {Pan}}\ and\ \bibinfo {author} {\bibfnamefont {S.}~\bibnamefont
  {Das~Sarma}},\ }\bibfield  {title} {\bibinfo {title} {Disorder effects on
  majorana zero modes: Kitaev chain versus semiconductor nanowire},\ }\href
  {https://doi.org/10.1103/PhysRevB.103.224505} {\bibfield  {journal} {\bibinfo
   {journal} {Phys. Rev. B}\ }\textbf {\bibinfo {volume} {103}},\ \bibinfo
  {pages} {224505} (\bibinfo {year} {2021})}\BibitemShut {NoStop}%
\bibitem [{\citenamefont {Pan}\ and\ \citenamefont
  {Das~Sarma}(2022)}]{DasSarma2022}%
  \BibitemOpen
  \bibfield  {author} {\bibinfo {author} {\bibfnamefont {H.}~\bibnamefont
  {Pan}}\ and\ \bibinfo {author} {\bibfnamefont {S.}~\bibnamefont
  {Das~Sarma}},\ }\bibfield  {title} {\bibinfo {title} {On-demand large
  conductance in trivial zero-bias tunneling peaks in majorana nanowires},\
  }\href {https://doi.org/10.1103/PhysRevB.105.115432} {\bibfield  {journal}
  {\bibinfo  {journal} {Phys. Rev. B}\ }\textbf {\bibinfo {volume} {105}},\
  \bibinfo {pages} {115432} (\bibinfo {year} {2022})}\BibitemShut {NoStop}%
\bibitem [{\citenamefont {Alicea}(2012)}]{Alicea2012}%
  \BibitemOpen
  \bibfield  {author} {\bibinfo {author} {\bibfnamefont {J.}~\bibnamefont
  {Alicea}},\ }\href@noop {} {\bibfield  {journal} {\bibinfo  {journal} {Rep.
  Prog. Phys. 75, 076501}\ } (\bibinfo {year} {2012})}\BibitemShut {NoStop}%
\bibitem [{\citenamefont {Dolcini}\ \emph {et~al.}(2015)\citenamefont
  {Dolcini}, \citenamefont {Houzet},\ and\ \citenamefont {Meyer}}]{Meyer2015}%
  \BibitemOpen
  \bibfield  {author} {\bibinfo {author} {\bibfnamefont {F.}~\bibnamefont
  {Dolcini}}, \bibinfo {author} {\bibfnamefont {M.}~\bibnamefont {Houzet}},\
  and\ \bibinfo {author} {\bibfnamefont {J.~S.}\ \bibnamefont {Meyer}},\
  }\bibfield  {title} {\bibinfo {title} {Topological josephson
  ${\ensuremath{\phi}}_{0}$ junctions},\ }\href
  {https://doi.org/10.1103/PhysRevB.92.035428} {\bibfield  {journal} {\bibinfo
  {journal} {Phys. Rev. B}\ }\textbf {\bibinfo {volume} {92}},\ \bibinfo
  {pages} {035428} (\bibinfo {year} {2015})}\BibitemShut {NoStop}%
\bibitem [{\citenamefont {Black-Schaffer}(2011)}]{Annica2011}%
  \BibitemOpen
  \bibfield  {author} {\bibinfo {author} {\bibfnamefont {A.~M.}\ \bibnamefont
  {Black-Schaffer}},\ }\bibfield  {title} {\bibinfo {title} {Self-consistent
  superconducting proximity effect at the quantum spin hall edge},\ }\href
  {https://doi.org/10.1103/PhysRevB.83.060504} {\bibfield  {journal} {\bibinfo
  {journal} {Phys. Rev. B}\ }\textbf {\bibinfo {volume} {83}},\ \bibinfo
  {pages} {060504} (\bibinfo {year} {2011})}\BibitemShut {NoStop}%
\bibitem [{\citenamefont {Fulde}\ and\ \citenamefont {Ferrell}(1964)}]{FFLO-1}%
  \BibitemOpen
  \bibfield  {author} {\bibinfo {author} {\bibfnamefont {P.}~\bibnamefont
  {Fulde}}\ and\ \bibinfo {author} {\bibfnamefont {R.~A.}\ \bibnamefont
  {Ferrell}},\ }\bibfield  {title} {\bibinfo {title} {Superconductivity in a
  strong spin-exchange field},\ }\href
  {https://doi.org/10.1103/PhysRev.135.A550} {\bibfield  {journal} {\bibinfo
  {journal} {Phys. Rev.}\ }\textbf {\bibinfo {volume} {135}},\ \bibinfo {pages}
  {A550} (\bibinfo {year} {1964})}\BibitemShut {NoStop}%
\bibitem [{\citenamefont {Larkin}\ and\ \citenamefont
  {Ovchinnikov}(1965)}]{FFLO-2}%
  \BibitemOpen
  \bibfield  {author} {\bibinfo {author} {\bibfnamefont {A.~I.}\ \bibnamefont
  {Larkin}}\ and\ \bibinfo {author} {\bibfnamefont {Y.~N.}\ \bibnamefont
  {Ovchinnikov}},\ }\bibfield  {title} {\bibinfo {title} {Inhomogeneous state
  of superconductors},\ }\href@noop {} {\bibfield  {journal} {\bibinfo
  {journal} {Soviet Phys. JETP}\ }\textbf {\bibinfo {volume} {20}},\ \bibinfo
  {pages} {762} (\bibinfo {year} {1965})}\BibitemShut {NoStop}%
\bibitem [{\citenamefont {Weithofer}\ and\ \citenamefont
  {Recher}(2013)}]{Weithofer2013}%
  \BibitemOpen
  \bibfield  {author} {\bibinfo {author} {\bibfnamefont {L.}~\bibnamefont
  {Weithofer}}\ and\ \bibinfo {author} {\bibfnamefont {P.}~\bibnamefont
  {Recher}},\ }\bibfield  {title} {\bibinfo {title} {Chiral majorana edge
  states in hgte quantum wells},\ }\href
  {https://doi.org/10.1088/1367-2630/15/8/085008} {\bibfield  {journal}
  {\bibinfo  {journal} {New Journal of Physics}\ }\textbf {\bibinfo {volume}
  {15}},\ \bibinfo {pages} {085008} (\bibinfo {year} {2013})}\BibitemShut
  {NoStop}%
\bibitem [{\citenamefont {De}\ \emph {et~al.}(2020)\citenamefont {De},
  \citenamefont {Khanna},\ and\ \citenamefont {Rao}}]{Suman2020}%
  \BibitemOpen
  \bibfield  {author} {\bibinfo {author} {\bibfnamefont {S.~J.}\ \bibnamefont
  {De}}, \bibinfo {author} {\bibfnamefont {U.}~\bibnamefont {Khanna}},\ and\
  \bibinfo {author} {\bibfnamefont {S.}~\bibnamefont {Rao}},\ }\bibfield
  {title} {\bibinfo {title} {Magnetic flux periodicity in second order
  topological superconductors},\ }\href
  {https://doi.org/10.1103/PhysRevB.101.125429} {\bibfield  {journal} {\bibinfo
   {journal} {Phys. Rev. B}\ }\textbf {\bibinfo {volume} {101}},\ \bibinfo
  {pages} {125429} (\bibinfo {year} {2020})}\BibitemShut {NoStop}%
\bibitem [{\citenamefont {Self}\ \emph {et~al.}(2020)\citenamefont {Self},
  \citenamefont {Rubio-Garc\'{\i}a}, \citenamefont {Garc\'{\i}a-Ripoll},\ and\
  \citenamefont {Pachos}}]{Pachos_2020}%
  \BibitemOpen
  \bibfield  {author} {\bibinfo {author} {\bibfnamefont {C.~N.}\ \bibnamefont
  {Self}}, \bibinfo {author} {\bibfnamefont {A.}~\bibnamefont
  {Rubio-Garc\'{\i}a}}, \bibinfo {author} {\bibfnamefont {J.~J.}\ \bibnamefont
  {Garc\'{\i}a-Ripoll}},\ and\ \bibinfo {author} {\bibfnamefont {J.~K.}\
  \bibnamefont {Pachos}},\ }\bibfield  {title} {\bibinfo {title} {Topological
  bulk states and their currents},\ }\href
  {https://doi.org/10.1103/PhysRevB.102.045424} {\bibfield  {journal} {\bibinfo
   {journal} {Phys. Rev. B}\ }\textbf {\bibinfo {volume} {102}},\ \bibinfo
  {pages} {045424} (\bibinfo {year} {2020})}\BibitemShut {NoStop}%
\bibitem [{\citenamefont {Zhang}\ \emph {et~al.}(2021)\citenamefont {Zhang},
  \citenamefont {Chen}, \citenamefont {Wu}, \citenamefont {Jiang},
  \citenamefont {Liu}, \citenamefont {Sun},\ and\ \citenamefont
  {Xie}}]{Sun_2021}%
  \BibitemOpen
  \bibfield  {author} {\bibinfo {author} {\bibfnamefont {Z.-Q.}\ \bibnamefont
  {Zhang}}, \bibinfo {author} {\bibfnamefont {C.-Z.}\ \bibnamefont {Chen}},
  \bibinfo {author} {\bibfnamefont {Y.}~\bibnamefont {Wu}}, \bibinfo {author}
  {\bibfnamefont {H.}~\bibnamefont {Jiang}}, \bibinfo {author} {\bibfnamefont
  {J.}~\bibnamefont {Liu}}, \bibinfo {author} {\bibfnamefont {Q.-f.}\
  \bibnamefont {Sun}},\ and\ \bibinfo {author} {\bibfnamefont {X.~C.}\
  \bibnamefont {Xie}},\ }\bibfield  {title} {\bibinfo {title} {Chiral interface
  states and related quantized transport in disordered chern insulators},\
  }\href {https://doi.org/10.1103/PhysRevB.103.075434} {\bibfield  {journal}
  {\bibinfo  {journal} {Phys. Rev. B}\ }\textbf {\bibinfo {volume} {103}},\
  \bibinfo {pages} {075434} (\bibinfo {year} {2021})}\BibitemShut {NoStop}%
\bibitem [{\citenamefont {Qu}\ \emph {et~al.}(2014)\citenamefont {Qu},
  \citenamefont {Gong},\ and\ \citenamefont {Zhang}}]{Chunlei_2014}%
  \BibitemOpen
  \bibfield  {author} {\bibinfo {author} {\bibfnamefont {C.}~\bibnamefont
  {Qu}}, \bibinfo {author} {\bibfnamefont {M.}~\bibnamefont {Gong}},\ and\
  \bibinfo {author} {\bibfnamefont {C.}~\bibnamefont {Zhang}},\ }\bibfield
  {title} {\bibinfo {title} {Fulde-ferrell-larkin-ovchinnikov or majorana
  superfluids: The fate of fermionic cold atoms in spin-orbit-coupled optical
  lattices},\ }\href {https://doi.org/10.1103/PhysRevA.89.053618} {\bibfield
  {journal} {\bibinfo  {journal} {Phys. Rev. A}\ }\textbf {\bibinfo {volume}
  {89}},\ \bibinfo {pages} {053618} (\bibinfo {year} {2014})}\BibitemShut
  {NoStop}%
\bibitem [{\citenamefont {Roy}\ and\ \citenamefont
  {Kumar}(2021)}]{Monalisa_2021}%
  \BibitemOpen
  \bibfield  {author} {\bibinfo {author} {\bibfnamefont {M.~S.}\ \bibnamefont
  {Roy}}\ and\ \bibinfo {author} {\bibfnamefont {M.}~\bibnamefont {Kumar}},\
  }\href@noop {} {\bibinfo {title} {Fulde-ferrel-larkin-ovchinnikov phase in
  one dimensional fermi gas with attractive interactions and transverse
  spin-orbit coupling}} (\bibinfo {year} {2021}),\ \Eprint
  {https://arxiv.org/abs/2108.03314} {arXiv:2108.03314 [cond-mat.quant-gas]}
  \BibitemShut {NoStop}%
\bibitem [{\citenamefont {Nakai}\ \emph {et~al.}(2021)\citenamefont {Nakai},
  \citenamefont {Nomura},\ and\ \citenamefont {Tanaka}}]{Tanaka_QAH_JJ}%
  \BibitemOpen
  \bibfield  {author} {\bibinfo {author} {\bibfnamefont {R.}~\bibnamefont
  {Nakai}}, \bibinfo {author} {\bibfnamefont {K.}~\bibnamefont {Nomura}},\ and\
  \bibinfo {author} {\bibfnamefont {Y.}~\bibnamefont {Tanaka}},\ }\bibfield
  {title} {\bibinfo {title} {Edge-induced pairing states in a {J}osephson
  junction through a spin-polarized quantum anomalous {H}all insulator},\
  }\href {https://doi.org/10.1103/PhysRevB.103.184509} {\bibfield  {journal}
  {\bibinfo  {journal} {Phys. Rev. B}\ }\textbf {\bibinfo {volume} {103}},\
  \bibinfo {pages} {184509} (\bibinfo {year} {2021})}\BibitemShut {NoStop}%
\bibitem [{\citenamefont {Tanaka}\ \emph {et~al.}(2007)\citenamefont {Tanaka},
  \citenamefont {Tanuma},\ and\ \citenamefont {Golubov}}]{Tanaka_OddFrequency}%
  \BibitemOpen
  \bibfield  {author} {\bibinfo {author} {\bibfnamefont {Y.}~\bibnamefont
  {Tanaka}}, \bibinfo {author} {\bibfnamefont {Y.}~\bibnamefont {Tanuma}},\
  and\ \bibinfo {author} {\bibfnamefont {A.~A.}\ \bibnamefont {Golubov}},\
  }\bibfield  {title} {\bibinfo {title} {Odd-frequency pairing in
  normal-metal/superconductor junctions},\ }\href
  {https://doi.org/10.1103/PhysRevB.76.054522} {\bibfield  {journal} {\bibinfo
  {journal} {Phys. Rev. B}\ }\textbf {\bibinfo {volume} {76}},\ \bibinfo
  {pages} {054522} (\bibinfo {year} {2007})}\BibitemShut {NoStop}%
\bibitem [{\citenamefont {Tanaka}\ \emph {et~al.}(2012)\citenamefont {Tanaka},
  \citenamefont {Sato},\ and\ \citenamefont {Nagaosa}}]{Tanaka_Review}%
  \BibitemOpen
  \bibfield  {author} {\bibinfo {author} {\bibfnamefont {Y.}~\bibnamefont
  {Tanaka}}, \bibinfo {author} {\bibfnamefont {M.}~\bibnamefont {Sato}},\ and\
  \bibinfo {author} {\bibfnamefont {N.}~\bibnamefont {Nagaosa}},\ }\bibfield
  {title} {\bibinfo {title} {Symmetry and topology in superconductors -
  odd-frequency pairing and edge states},\ }\href
  {https://doi.org/10.1143/JPSJ.81.011013} {\bibfield  {journal} {\bibinfo
  {journal} {Journal of the Physical Society of Japan}\ }\textbf {\bibinfo
  {volume} {81}},\ \bibinfo {pages} {011013} (\bibinfo {year}
  {2012})}\BibitemShut {NoStop}%
\bibitem [{\citenamefont {Park}\ \emph {et~al.}(2017)\citenamefont {Park},
  \citenamefont {Yang}, \citenamefont {Kim},\ and\ \citenamefont
  {Gilbert}}]{MoonJip_2017}%
  \BibitemOpen
  \bibfield  {author} {\bibinfo {author} {\bibfnamefont {M.~J.}\ \bibnamefont
  {Park}}, \bibinfo {author} {\bibfnamefont {J.}~\bibnamefont {Yang}}, \bibinfo
  {author} {\bibfnamefont {Y.}~\bibnamefont {Kim}},\ and\ \bibinfo {author}
  {\bibfnamefont {M.~J.}\ \bibnamefont {Gilbert}},\ }\bibfield  {title}
  {\bibinfo {title} {Fulde-ferrell states in inverse proximity-coupled
  magnetically doped topological heterostructures},\ }\href
  {https://doi.org/10.1103/PhysRevB.96.064518} {\bibfield  {journal} {\bibinfo
  {journal} {Phys. Rev. B}\ }\textbf {\bibinfo {volume} {96}},\ \bibinfo
  {pages} {064518} (\bibinfo {year} {2017})}\BibitemShut {NoStop}%
\end{thebibliography}%

\end{document}